\magnification=1200
\baselineskip=7mm
\def\sect#1{\vskip 6mm \noi {\bf #1} \vskip 4mm}
\def\section#1{\vskip 6mm \noi {\bf #1} \vskip 4mm}
\def\subsection#1{\vskip 6mm \noi {\bf #1} \vskip 4mm}
\def\subsect#1{\vskip 6mm \noi {\bf #1} \vskip 4mm}
\def\Ha{H$\alpha$}
\def\v{\vskip 2mm}
\def\noi{\noindent}
\def\r{\hangindent=1pc  \noindent}
\def\cen{\centerline} 
\def\endpage{\vfil\break}
\def\kms{km s$^{-1}$}
\def\vlsr{$V_{\rm LSR}$}
\def\Vrot{V_{\rm rot}}
\def\Msun{M_{\odot \hskip-5.2pt \bullet}}
\def\Deg{^\circ}
\def\deg{$^\circ$}
\def\ta{$T_{\rm a}$}
\def\Ta{T_{\rm a}}
\def\co{CO ($J=1-0$) }

\topskip 15mm

\cen{\bf Nuclear Rotation Curves of Galaxies in the CO Line Emission}

\topskip 0mm 
\vskip10mm
\cen{Yoshiaki SOFUE,  Yoshinori TUTUI, Mareki HONMA and Akihiko TOMITA}

\cen{\it Institute of Astronomy, University of Tokyo, 2-21-1 Osawa, Mitaka, Tokyo 181}

\cen{\it E-mail: sofue@mtk.ioa.s.u-tokyo.ac.jp}

\centerline{(1997 May)}
\vskip 10mm

\centerline{Abstract} 

We have obtained high-resolution 
position-velocity (PV) diagrams along the major
axes of the central regions of nearby galaxies in the CO-line
emission using the Nobeyama 45-m telescope and the Millimeter Array.
Nuclear rotation curves for 14 galaxies have been derived 
based on the PV diagrams using the envelope-tracing method.
The nuclear rotation curves for most of the galaxies show a 
steep rise   within a few hundred pc,
which indicates a high-density concentration of mass. 


\sect{1. Introduction}

Rotation curves (RC) of galaxies   have been supposed to generally show
a rigid-body rise in the central few kpc region, followed
by a flat rotation in the disk and outer regions 
(Rubin et al. 1980, 1982; Mathewson et al 1996;   Persic et al. 1995).
Although the HI rotation curves are most useful for investigating the
dark halo and total mass for its extended distribution in the outer 
region (e.g., Kent 1987), the HI gas is deficient in the central regions.
Therefore, HI rotation curves may be not accurate enough to discuss the
nuclear rotation.
The current optical rotation curves, which were obtained 
primarily for the study of dark halo in the outermost regions,
had been over-exposed in the central regions, and may not indicate
the true nuclear rotation within the bulges. 

It is well known that the detailed rotation curve for
our Milky Way Galaxy shows a very steep rise in the central few
hundred pc (e.g., Clemens 1985).
A simple question arises, if the true rotation curves of galaxies,
particularly for the innermost regions, would be more like that 
for our Galaxy.
In order to clarify this question, we have proposed to
use the CO line data for the central regions, where the
molecular gas dominates over the HI gas (Sofue et al 1994; Honma et al 1995).
We have recently used CO-line data for the central parts
of nearby galaxies, and obtained combined rotation curves with
the outer HI and optical curves (Sofue 1996, 1997:Paper I and II).
We have shown that the nuclear CO rotation curves generally 
rise sharply in the central few hundred parsecs than
those derived from HI and optical observations alone.
More recently, Rubin et al (1997) have also 
shown that the central rotation curves
rise steeply within a few hundred pc, indicating a massive
rotating disk around the nucleus, 
from their \Ha\ spectroscopy of Virgo-cluster spirals.

In this paper, we present the result of high-resolution
CO-line observations along the major axes of
the central regions of nearby galaxies using the Nobeyama 
45-m telescope and the NMA (Nobeyama Millimeter Array). 
We aim to provide basic data for the study of nuclear rotation
curves, as well as to clarify whether the nuclear steep rise
is indeed a universal   characteristics for spiral galaxies.

\sect{2. Observations}

\subsect{2.1. Observations with the 45-m Telescope}
 
Observations of the \co\ line with the 45-m telescope 
were made on 1996 March 16 -- 20 (NGC 2903, 3521, 4631, 5055, 7331), 
and December 18 -- 20 (NGC 598, 1003, 2403, 
3198,  3953, 4096, 5457 (M101), 6674, 
and UGC 2855).
The antenna had a HPBW of $15''$ at the CO line frequency, 
and the aperture and  main-beam efficiencies were  
$\eta_{\rm a}=0.35$ and $\eta_{\rm mb}=0.50$, respectively.
We used two  SIS (superconductor-insulator-superconductor)
receivers with orthogonal polarization,
which were combined with 2048-channel acousto-optical spectrometers.
The total channel number corresponds to a frequency width of 250 MHz,
and, therefore, to a velocity coverage in the rest frame at the galaxy 
of 650  \kms with a resolution of 650/2048 \kms. 
The center frequency was tuned to coincide with the systemic velocity of
the galaxies.

We scanned the major axis of each galaxy at a spacing of $7''.5$ for the
central $\pm 30''$ region, and at $15''$ for the outer regions. 
After combining every 32 channels in order to increase the signal-to-noise 
ratio, we obtained spectra with a velocity resolution of 10.2 \kms. 
The calibration of the line intensity was made using an absorbing chopper 
in front of the receiver, yielding an antenna temperature (\ta), corrected 
for both the atmospheric and antenna ohmic losses.
We used a multiple-on and off switching mode, and the on-source total 
integration time per data point was typically fifteen minutes.
The system noise temperature (SSB) was 500 to 800 K.
After flagging and subtraction of baselines, the spectra were
smoothed to a velocity resolution of 10 \kms (32 channels).
The rms noise of the resultant spectra 
at a velocity resolution of 10 \kms\ was typically 20 mK in \ta.
The pointing of the antenna was corrected and checked by observing 
nearby SiO maser sources at 43 GHz every 1 to 1.5 hours, 
and was typically within $\pm 3''$. 
 
\subsect{2.1. Observations with the Nobeyama Millimeter Array}
 
High-resolution interferometer observations 
were obtained by using  the  Nobeyama Millimeter 
Array (NMA) in the C configuration on 
1994 December 1--3 (NGC 4501, NGC 4527, NGC 4303),
December 15--17 (NGC 4569), and in the D-configuration on
 1995 March 15 -- 17, and March 29-31. 
In this paper we present the results for NGC 4303 and NGC 4569,
for which good PV diagrams have been obtained.
The flux and phase calibrations were done by oberving the
nearby radio source 3C 273, which had a flux density of 22 Jy
at the observing frequency.
A 1024-channel FX system (a fast-Fourier-transform
spectro-correlator) was used for the  spectroscopic data
acquisition  with a total bandwidth of 320 MHz (831 \kms).
The data were then averaged in 16 bins of original frequency channels
resulting in 64 channels with a frequency (velocity) resolution of 5 MHz
(13.0 \kms). 
The cleaned maps were then combined
into a cube of intensity data in the (RA, Dec, \vlsr) space.
Using these data cube, we obtained position-velocity diagrams along
a constant declination across the nucleus for NGC4303, approximately
parallel to the major axis (PA 105\deg), and along a constant RA for 
NGC4569 (major axis at PA 23\deg).

\subsection{2.3. Selection of Objects}

Observed positions and systemic velocities of the program 
galaxies are listed in Table 1.
Table 2 list the adopted parameters for these galaxies.
The galaxies were so chosen that 
(a) the angular size is large enough in order to obtain 
a sufficiently high linear-scale resolution in the central regions; 
(b) the disk is mildly tilted in order for an accurate
correction for the inclination to derive the rotation velocity;  
(c) no high-sensitivity data have been obtained yet with the 
45-m telescope or the NMA; and 
(d) the CO-line emission is sufficiently strong or 
IRAS 60 and 100 $\mu$m fluxes are higher than several Jy.
For the selection of objects,  we have referred to the 
published CO data from the Five College 14-m telescope
(Young et al 1995; Kenney and Young 1988), and the
NRO 45-m telescope (Nishiyama 1995).

\v
\cen{ -- Tables 1 and 2 --}

\section{3. Results}\v

\subsection{3.1. Position-Velocity Diagrams}
  
The obtained data are presented in Fig. 1 in the form of 
position-velocity (PV) diagrams.
We also show intensity profiles along the major axis and averaged velocity
profiles, which have been obtained by averaging the PV diagrams in the
velocity and position directions, respectively.
Below, we describe individual galaxies. Velocities used in the
description are corrected for the inclinations.

\v
\cen{--Fig. 1 --}
\v

NGC 598 (M33): This nearby Sc galaxy is a member of the local
group, showing diffuse, rather amorphous, spiral arms, 
and the bulge is small in size and luminosity. 
The PV diagram has been obtained along PA=54\deg, about 30\deg
different from the true PA at 23\deg.
The CO emission is weak, and no rotating disk is visible 
in the central  30$''$ (110 pc).
Therefore, no rotation curve was obtained for this galaxy.

NGC 1003: This is a  tilted, two-armed spiral. The CO-line
intensity is asymmetric with respect to the nucleus. 
The PV diagram indicates a steep rise of rotation near the nucleus
in the negative velocity side, while the opposite side is
only weakly detected.
The gradient of rotation rise is as large as $\sim 250$ \kms / 10$''$
(460 pc).
 
NGC 2403: This is an Sc galaxy with open spiral arms,
morphologically similar to M33.
The nuclear PV diagram shows a steep rise, reaching $\Vrot \sim 170$ \kms
within the central $10''$ (160 pc).
Although the morphology is similar, the nuclear
rotation is different from that for M33.

UGC 2855: This is an SBc type galaxy with high optical surface brightness,
associated with a companion (UGC 2866) at a distance of 
10$'$ (84 kpc) to the east.
The CO intensity as strong as $\Ta \sim 0.45$ K has been observed
in the central part.
The PV diagram shows a very steep rise to $\Vrot \sim 200$ \kms within the 
central $\sim 7''$ (1 kpc), indicating a rapidly rotating disk. 
The negative velocity peak of this central disk attains a maximum 
velocity as high as 200 \kms / sin $i$ = 220 \kms within the 
central few arcseconds.
Besides this steeply increasing component, the PV diagram indicates a
rigid-body like ridge with a velocity gradient of 170 \kms / 30$''$ (4 kpc),
followed by a flat rotation part at 200 \kms at $>30''$.

NGC 2903: The PV diagram shows a steep rise within the central $7''$ (280 pc)
to $\Vrot \sim 160$ \kms, and has a maximum at $\sim 15''$ (600 pc)
at $\Vrot \sim 250$ \kms.
The main ridge of the PV diagram in the southern part has a peculiar step 
at 1$'$ (2.4 kpc) radius, with a flat part of about 170 \kms\ till 1$'$,
and, then, steps up to a flat part at 200 \kms. 
The outer rotation from HI, however, is very flat until the edge.

NGC 3034 (M82): This peculiar edge-on galaxy has been extensively studied
in the CO line. The rotation curve is known for its steep rise near the
center, and is declining in a Keplerian fashion, suggesting a truncated
outer mass by a close encounter with the parent massive galaxy, M81
(Sofue et al. 1992). Our data show the nuclear rise and high-density
gaseous torus rotating at about 120 \kms, consistent
with the earlier observations. 
We show this galaxy in order to demonstrate that the nuclear rise of
rotation is common even to such a dwarf and peculiar
galaxy as M82 in a high starburst activity.

NGC 3198: The PV diagram appears to consist of two components:
a steeply rising ridge, attaining a high-velocity peak within 20$''$,
and a rigidly rising component with a milder slope.
The first rotation peak of $V \sim 170$ \kms occurs at $r=23''$ (1100 pc). 
The rigid-body component has a gradient of 110 \kms in 30$''$ (1.4 kpc),
probably due to a ring or arms of a radius of about 30$''$.

NGC 3521: The inner rotation velocity increases steeply, and attains
a small and sharp central peak at 10$''$  (520 pc) of 210 \kms, 
followed by a dip at 20$''$ (1 kpc). It then increases slowly until
a broad maximum at $1'.5$ (4.7 kpc).
However, the very center shows very weak emission, indicating
either little CO gas in the center, or a very high velocity
dispersion. 
The outer HI RC gradually decreases until the edge.

NGC 4096: This is an Sc galaxy with north-south asymmetry in
the optical brightness distribution.
The central CO distribution is also asymmetric, with the
northern half stronger than the south.
The PV diagram indicates a rotating ring of radius $\sim 20''$  with a
rotating velocity of $V_i \sim 85$ \kms.
The CO rotation is followed by a flat part at a velocity of
$\sim 185$ \kms.
The nuclear few arcsec region shows no significant CO emission. 

NGC 4303: This is a nearly face-on galaxy of Sc type, having a sign
of bar. The CO intensity sharply peaks near the nucleus, and shows 
a high concentration within a radius of 15$'' $(0.8 kpc). 
The rotation velocity increases very steeply within 0.5 kpc
to about 120 \kms.

NGC 4569: This Sab type galaxy shows also a high concentration of CO
intensity near the center. 
The velocity increases very sharply within the central 5$''$
(300 pc) to about 200 \kms.

NGC 4631: This is an almost edge-on amorphous Sc galaxy, rich in
CO gas. 
The rotation velocity increases in a rigid-body fashion, as slowly as
$\sim 100$ \kms/1 kpc.
This galaxy is one of the exceptions, which show rigid-body
rotation.

NGC 5055: The PV diagram shows a very steep rise within $\sim 6''$ (210 pc)to 
a velocity of $V_i \sim 200$ \kms. Then, the rotation
velocity is almost constant till $\pm 40''$ (1.4 kpc).

NGC 5457 (M101): This is a nearly face-on Sc galaxy with  $i \sim 18 \Deg$. 
CO emission has been detected near the systemic velocity (241 \kms), with
a velocity width of $\sim 70$ \kms, which corresponds to 230 \kms
within the disk plane.
The northern half shows stronger emission with a negative velocity, 
while the southern half is not clearly detected.
However, if two weak emission peaks in the north at $V_{\rm hel} \sim 170$ \kms
and in the south at $V_{hel}\sim 350$ \kms are due to the rotation,
the rotation velocity is estimated to be as high as $\sim 290 $ \kms
within the central $\sim 10''$ (350 pc).
For the poor data and low inclination, no rotation curve has been obtained. 

NGC 6647: This is a bar-ringed galaxy, and weak CO emission has
been detected. 
The obtained PV diagram is not of high-quality, but can be
used to trace rotation velocities, indicating a steep rise
of rotation up to 250 \kms within a few hundred pc of the
center.

NGC 7331: The PV diagram shows a rigid-body increase, corresponding to
a ring-like distribution of gas at a radius 30$''$ (2 kpc) rotating
at $\Vrot= 250$ \kms. The rotation is then flat to $\pm 2'$ (8 kpc).
The central 20$''$ (1.4 kpc) region shows weak CO emission. 
The averaged velocity profile shows a typical double-horn
property, characteristic to a rotation disk with a constant velocity.

\subsection{3.2.  Nuclear Rotation Curves}

We adopt the envelope-tracing method (Sofue 1996) to 
derive rotation curves, which  uses the loci of terminal  
velocity in position-velocity (PV) diagrams.
We define the terminal velocity by a velocity at which
the intensity becomes equal to
$$
I_{\rm t}=[(0.2 I_{\rm max})^2+I_{\rm lc}^2]^{1/2} \eqno(1)
$$
on the PV diagrams.
Here,  $I_{\rm max}$ and $I_{\rm lc}$ are 
the maximum intensity and intensity corresponding to the lowest 
contour level, respectively.
This equation defines a 20\% level of the intensity profile at a fixed 
position, $I_{\rm t}\simeq 0.2 \times I_{\rm max}$,
if the signal-to-noise ratio is sufficiently high.
If the intensity is not high enough, the equation gives
$I_{\rm t}\simeq I_{\rm lc}$, which approximately defines the loci 
along the lowest contour level ($\sim 3 \times$ rms noise).
The terminal velocity is then corrected for the 
velocity dispersion of the interstellar gas
($\sigma_{\rm ISM}$) and the velocity resolution of observations 
($\sigma_{\rm obs}$) as 
$$
V_{\rm t}^0=V_{\rm t}-(\sigma_{\rm obs}^2 + \sigma_{\rm ISM}^2)^{1/2}.
\eqno(2)
$$ 
Small-scale structures due to clumpy ISM and clouds, and partly 
due to noise in the observations, are smoothed by eye estimates. 
The asymmetry with respect to the center of the PV diagram has
been often observed. In such a case with asymmetry, we have 
averaged the measured velocities in both sides of the nucleus.

The rotation velocity is finally obtained by 
$$ 
V_{\rm rot}=V_{\rm t}^0/{\rm sin}~i, \eqno(3)
$$
where $i$ is the inclination angle of the disk plane.
The accuracy of determining the terminal velocity, and therefore 
the accuracy of the obtained rotation curve, is typically 
$\pm 10 $/sin $i$ \kms.
However, in some cases the rotation curves in both sides of the
nucleus were not symmetric, and we have averaged the values
in both sides. 
This often caused errors amounting to 15 to 20 \kms,
larger than the error in the tracing
of PV diagram itself.

Fig. 2 shows the thus obtained rotation curves corrected for
the inclination for 14 galaxies, for which PV diagrams with 
sufficient signal-to-noise ratio have been obtained.
Fig. 3 shows all curves plotted in the same scales both in radius and velocity.
The horizontal and vertical bars of a  cross attached to each
curve indicate the FWHM of the telescope beam (linear length corresponding
to $15''$) and the typical velocity error as described above, respectively.
The radial distances were calculated by
using the distances listed in Table 1, which were
estimated by applying the B-band Tully-Fisher relation
(Pierce and Tully 1992) using the data from the
Reference Catalogue of Bright Galaxies (RC3; de Vaucouleurs
et al 1991). 
$$ {M}_{B}^{b,i} = -7.48(\rm log {W}_{\rm i~ 20} - 2.50)
- 19.55. $$
Here, ${W}_{\rm i~ 20}$ and ${M}_{B}^{b,i}$ are, respectively, 
the HI linewidth measured at 20\% level 
and B-band total absolute magnitude corrected for Galactic and internal 
extinction and for redshift taken from RC3. 

\v
\centerline{-- Fig. 2, 3 --}
\v

\sect{4. Discussion}

\subsect{4.1. Distribution of Molecular Gas}

The intensity distribution of the CO-line emission along the
major axis for the observed galaxies show more or less an 
asymmetry with respect to the nuclei. 
Such positional asymmetry and displacement 
of the CO intensity  has been also observed 
in the Milky Way (e.g.,  Bally et al 1987), and therefore, 
may be a common phenomenon in the central molecular disks of spiral 
galaxies.

However, if we adopt the kinematical centers of the
observed CO PV diagrams, the rotation velocities, as defined by 
the envelope-tracing method, are symmetric,
in spite of the significant asymmetry of the intensity. 
This would be reasonable, if the mass of molecular gas is not large
to affect the background gravitational potential of the inner disk and
bulge of the galaxy. 
This is consistent with the fact that the mass fraction of the molecular 
gas within a few hundred pc of a galaxy is only several percent 
of the dynamical mass (Sofue 1995), if we adopt the new conversion factor which
is a strong function of the metallicity (Arimoto et al 1996).

\subsection{4.2. Steep Rise of Rotation Curve}

Fig. 2 and 3 show that the rotation curves generally rise steeply
within a radius smaller than the beam width of the CO
observations, typically within a few hundred parsecs.
This agrees with our earlier study of other galaxies (Sofue 1996,1997),
and with the recent \Ha\ study of nuclear rotating disks in
Virgo-cluster galaxies (Rubin et al 1997).
Although the central rising parts for some galaxies mimic 
rigid rotation, they may be mostly due to a lack of resolution. 
An exception is NGC4631: In spite of the sufficient
resolution, this galaxy shows a gradually rising rotation
in a rigid-body fashion.
NGC 7331 may be another case which shows an apparent
rigid-body rise, while the CO emission of the very central region
is too weak to prove this property. 
Such an apparently rigid-body behavior of PV diagram is 
often caused by a lack of molecular gas in the central region:
A molecular ring in a highly-tilted galaxy would result in 
a rigid-body-like ridge on a PV diagram. 
Therefore, we may conclude that the steep rise, usually within 
a radius not resolved by the observing beam, is a general 
characteristics of rotation in the central regions of the observed 
galaxies.

In Paper I and II,  we have shown that the steeply rising rotation curves 
can be fitted by a model with a mass distribution with a high central
concentration, higher than that corresponding to the normal bulge
component (Sofue 1996). Particularly, such a steep rise within the central 100 
parsecs as obtained by high-resolution interferometer observations
(e.g., NGC 4303, NGC 4569) indicates the existence of a compact 
nuclear mass of a 100 to 150 pc radius and  a mass of several $10^9\Msun$. 
Since the mass within the central 100 pc region is dominated by that
of the bulge, such mass concentration may be deeply coupled with 
the dynamical structure and evolution of the central bulges. 
A detailed deconvolution of the rotation curves and comparison with
optical and infrared surface photometry will be given in a separate
paper.

\subsection{4.3. Small Effect of a Bar?}
 
If the molecular gas is in a non-circular motion such as due to a bar, 
it is not straightforward to use them to derive the mass.
In Paper I and II, we have shown that the 
probability of looking at a bar parrallely 
is much smaller compared to that of looking at it 
perpendicularly or at a large angle, and therefore, such 
non-circular motion, if it is dominant, 
would result in a more number of galaxies showing
rigid-body rotation than those showing a steep rise.
However, our study, together with those in Paper I and II, 
shows that most galaxies have a steep nuclear rise.
This fact indicates that the molecular gas is more likely to be
rotating circularly, than to be in a bar-shocked non-circular motion.
We finally mention that the mass distribution derived 
by assuming circular rotation
have shown a good agreement with their surface photometry 
at radii of 1 to 5 kpc (e.g., Kent 1987), where
the effect of bars would be most efficient.  

\v\v

The authors thank the staff of Nobeyama Radio Observatory for their
help during the observations.
This work has been performed as a common-use program of NRO.
 
\vskip 5mm
\noindent{\bf References} 
\vskip 5mm
 
\r Arimoto, N., Sofue, Y., Tsujimoto, T. 1996, PASJ 48, 175.

\r{Bally, J., Stark, A.A., Wilson, R.W., and Henkel, C. 1987, ApJ Suppl 65, 13.}

\r Clemens  D. P. 1985, ApJ 295,  422 

\r de Vaucouleurs G., de Vaucouleurs A., Corwin  H. G. Jr., et al.,
   1991, in {\it  Third Reference Catalogue of Bright Galaxies}
   (New York: Springer Verlag)

\r Honma, M., Sofue, Y., Arimoto, N. 1995, AA 304, 1
 
\r Kenney  J.,  Young  S. J. 1988, ApJS 66,  261  

\r Kent  S. M. 1987, AJ 93,  816         
 
\r Nishiyama  K. 1995 PhD Thesis  University of Tokyo.

\r Pierce M. J., Tully R. B. 1992, ApJ, 387, 47
 
\r Rubin  V. C.,  Ford  W. K.,  Thonnard  N. 1980,  ApJ  238,  471 

\r Rubin  V. C.,  Ford  W. K.,  Thonnard  N. 1982,  ApJ  261,  439 

\r Rubin V. C., Kenney, J. D. P., and Young, J. S.  1997 AJ 113, 4.

\r Sofue, Y. 1995, PASJ, 47. 527.

\r Sofue  Y. 1996,  ApJ  458,  120 (Paper I). 

\r Sofue, Y. 1997, PASJ 49, 17 (Paper II).

\r Sofue, Y., Honma, M., Arimoto, N. 1994, AA 296, 33-44.

\r Sofue, Y., Reuter, H.-P., Krause, M., Wielebinski, R., and Nakai, N.
1992, ApJ 395, 126.

\r Sofue  Y.,  Honma  M.,  Arimoto  N. 1994,  A\&A 296,  33 
  
\r Young  J.S.,  Xie  S.,  Tacconi  L.,  Knezek  P.,  Vicuso  P., 
Tacconi-Garman  L.,  Scoville  N.,  Schneider  S.,  et al.  1995, ApJS 98,  219
 
\endpage

\settabs 7 \columns
\def\d{\dotfill}
\def\v{\vskip 4mm}

\noindent Table 1: Parametes for the observed galaxies.

\vskip 2mm \hrule \vskip 2mm

\+ Galaxy & Type & RA$_{1950}$ & Decl$_{1950}$ 
& $V_{\rm sys, (hel~ or~ lsr) }$&Tele/BW &Obs. Date   \cr

\+  &&(h~~ m~~ s) & ($~~^\circ ~~~ ' ~~~ ''$) & (km/s) &  $~~~~~~('')$ \cr

\vskip 2mm \hrule \vskip 2mm

\+ NGC 598 & Sc\d&   01 31 01.67 & 30 24 15.0 &-175(hel)&45m/15 &Dec-1996 \cr
\+ NGC 1003& Scd\d& 02 36 06.12  & 40 39 28.0&620(hel) &45m/15 &Dec-1996 \cr
\+ NGC 2403& Sc\d &  07 32 05.50 & 65 42 40.0 &100(lsr)&45m/15 &Dec-1996\cr 
\+ UGC 2855& SBc\d &03 43 15.70  & 69 58 46.0 &1200(hel)&45m/15 &Dec-1996\cr 
\+ NGC 2903 &Sc\d &09 29 20.30   & 21 43 23.9 &550(lsr) & 45m/15&Mar-1996\cr
\+ NGC 3034  &Ir\d&   09 51 43.48 & 69 55 00.8 &200(lsr) & 45m/15& Dec-1996 \cr
\+ NGC 3198 &SBc\d &10 16 51.94 & 45 48 06.0 &665(lsr)& 45m/15& Dec-1996\cr
\+ NGC 3521 &Sbc\d&11 03 15.10  & 00 13 58.0 &800(lsr)& 45m/15&  Mar-1996\cr   
\+ NGC 4096 &Sc\d&12 03 28.48   & 47 45 20.0 &565(hel)& 45m/15 & Dec-1996\cr   
\+ NGC 4303 &Sc\d& 12 19 21.40  & 04 44 58.0  &1560(hel)
	&  NMA/$9.2^\times 5.7 _{{\rm PA}=-14^\circ.9}$,  Dec-1994\cr
\+ NGC 4569 &Sab\d&13 26 18.00  & 12 34 18.7 &240(hel) 
	& NMA/$ 9.9^\times  4.8 _{{\rm PA}=161^\circ.8}$,  Dec-1994  \cr
\+ NGC 4631 &ScIr\d& 12 39 39.70 & 32 48 48.0 &600(lsr)& 45m/15 &Dec-1992 \cr
\+ NGC 5055 &Sbc\d&  13 13 34.90 & 42 17 34.0  &555(lsr)&45m/15 & Mar-1996\cr 
\+ NGC 5457 &Sc\d &14 01 26.31   & 54 35 17.9& 240(hel) &45m/15 & Dec-1996\cr 
\+ NGC 6674 &SBb\d &  18 36 31.00& 25 19 55.0 &340(hel)&45m/15 &Dec-1996\cr 
\+ NGC 7331 &Sbc\d & 22 34 46.66 & 34 09 20.9 &825(lsr)& 45m/15 &Mar-1996\cr 

\vskip 2mm \hrule 
\endpage

\settabs 5 \columns
\def\d{\dotfill}
\def\v{\vskip 4mm}

\noindent Table 1: Observed galaxies.
\vskip 2mm \hrule \vskip 2mm
\+ Galaxy & Type & Inclination$^\dagger$ & P.A.$^\dagger$ 
&  Distance$^*$ &Linear Reso.  \cr
\+ & &(deg) &(deg) & (Mpc) &(pc/15$''$) \cr
\vskip 2mm \hrule \vskip 2mm

\+ NGC 598 (M33)  & Sc &  54 & 23(Obs:54)& 0.79 & 55.3 \cr

\+ NGC 1003 & Scd & 66 & 97 & 9.52 &692 \cr


\+ NGC 2403 & Sc & 60& 127 & 3.32& 241 \cr


\+ UGC 2855 & SBc & 61 & 105 & 28.77 &2092 \cr

\+ NGC 2903 & Sc & 66 & 17&  8.20 &596 \cr


\+ NGC 3034 (M82) & Ir & $\sim 90$ & 80 & 3.25 &236\cr

\+ NGC 3198 & SBc & 70 & 35 & 9.96 &724\cr

\+ NGC 3521 &  Sbc & 75& 163& 10.82 &787 \cr


\+ NGC 4096 &Sc & 73 & 20  & 12.22 &889\cr

\+ NGC 4303 & Sc & 27& 105(Obs:90) & 11.17 &$498\times310$ \cr

\+ NGC 4569 & Sab & 63 & 23(Obs:0) & 10.65 & $510\times250$ \cr

\+ NGC 4631 &  ScIr & 84& 86&  4.30 &313 \cr

\+ NGC 5055 & Sbc &  55& 105 & 9.82 &714   \cr

\+ NGC 5457(M101)   &  Sc & 18 & 38&  9.34 &679 \cr

\+ NGC 6674 &SBb &55 & 143 & 42.62 &3100\cr

\+ NGC 7331&      Sbc &  75& 168 & 13.55 &985  \cr

\v
\hrule
\v

$\dagger$ Inclination and position angle are taken 
from the ``Third Reference Catalogue of Bright Galaxies (RC3)"
(de Vaucouleurs et al.,1991).

* The distances have been derived by applying the Tully-Fisher
relation, as deseived in the text.

\endpage

\r Figure Captions

\v 
\r Fig. 1. Position-velocity (PV) diagrams in the CO ($J=2-1$) line emission
along the major axes of nearby galaxies. The intensity scales are
in $T_a^*$ in K, and contour levels are indicated by the small tables.
The left-upper panels show the distribution of CO intensity along
the major axis, and the right-lower panels are averaged line profiles. 
 
\v
\r Fig. 2. Nuclear rotation curves for 14 galaxies,
as derived by using the CO-line PV diagrams in Fig. 1.
The velocities have been corrected for inclination, and the 
radii have been calculated corresponding to the distances given in Table 1.  
Typical errors in velocity (vertical bar) and resolution 
(horizontal bar) are given by the crosses attached to individual curves.

\v
\r Fig. 3. The same as Fig. 2, but all curves are plotted in
the same linear and velocity scales.

\bye